\documentclass[12pt]{iopart}

\newcommand{\pT}{\ensuremath{p_{\mathrm{t}}}}
\def\ave#1{\langle {#1} \rangle}

\begin{document}

\title[Proton-proton physics in ALICE]{Proton-proton physics in ALICE}
\author{Tapan K. Nayak\footnote[7]{On leave from Variable Energy Cyclotron Centre, Kolkata, India.} for the ALICE Collaboration}
\address{CERN, CH-1211, Geneva 23, Switzerland}
\ead{Tapan.Nayak@cern.ch}

\begin{abstract}
The ALICE experiment has several unique features which makes it an important
contributor to proton-proton physics at the LHC, in addition to its specific
design goal of studying the physics of strongly interacting matter in heavy-ion
collisions. The unique capabilities include its 
low transverse momentum (\pT) acceptance, excellent vertexing, particle identification over
a broad \pT~range and jet reconstruction. In this report, a brief
review of ALICE capabilities is given for studying bulk properties of produced particles which
characterize the underlying events, and the physics of heavy-flavour, quarkonia,
photons, di-leptons and jets.
\end{abstract}

\section{Introduction}
The Large Hadron Collider (LHC), under construction at CERN, is designed to deliver colliding 
proton-proton beams at center-of-mass energy of 14~TeV and lead-lead beams at
$\sqrt{s_{\rm NN}} = 5.5$~TeV. Collisions at these previously unexplored energies
offer outstanding opportunities for new physics in compliance with the Standard Model
and beyond. For the \mbox{p--p} collisions in particular,
in order to fully exploit the enormous physics potential, it is important to have a complete
understanding of the reaction mechanism including the interplay of non-perturbative to 
perturbative phenomena \cite{leshouches2005}. 
This is possible 
with the measurements of particle multiplicity, spectra, strangeness,
heavy-flavour, photon and di-lepton production, particle correlations, quarkonia physics,
bottom and charm quark cross sections as well as physics of jets \cite{aliceppr1,aliceppr2}.
Study of the underlying event structures of high multiplicity events and 
exploration of non-perturbative strong-coupling phenomena
related to confinement and hadronic structure at low Bjorken-{\it x} 
values will be of importance. 
Typical values of the cross sections \cite{leshouches2005,kaidalov} for \mbox{p--p} collisions 
at 14~TeV are 102~mb for the total, 76~mb for non-diffractive inelastic processes, 12~mb for single 
diffractive processes and 1~mb for bottom-quark production. The main contribution will be in
the low transverse-momentum (\pT) region for which the ALICE detector is optimized.
The LHC will also give access to data above the ``knee'' of
the cosmic-ray energy distribution for the first time. In addition to these topics, 
the \mbox{p--p} programme of ALICE forms an important benchmark for heavy-ion physics.

\section {ALICE experiment} 
The ALICE experiment is specifically designed to 
study the physics of strongly interacting matter and the quark-gluon plasma in heavy-ion
collisions. The low material budget and low magnetic field of the central barrel
makes the experiment sensitive to low-\pT~ particles (down to 100~MeV/{\it c} depending on the
particle type).
The Inner Tracking System (ITS), consisting of six layers of silicon detectors,
has excellent vertexing capabilities. 
The combination of ITS and a large Time Projection Chamber (TPC) provides powerful tracking
with excellent momentum resolutions from 100~MeV/{\it c} to $\sim$5\% at 100~GeV/{\it c}.
The particle identification of hadrons is provided with the inclusion of a 
Time-Of-Flight (TOF) system and 
a single arm Cherenkov detector (RICH). The central barrel is also equipped with a 
Transition Radiation Detector (TRD) 
for electron identification and a single arm photon calorimeter (PHOS). 
The new electromagnetic calorimeter (EMCAL) will improve 
ALICE capability for measurement of high energy jets.
The experimental 
capabilities are enhanced by a forward muon arm, a Forward Multiplicity Detector (FMD),
and a Photon Multiplicity Detector (PMD). The experiment is equipped with Zero Degree
Calorimeters (ZDC) and scintillator detectors for trigger and timing (V0, T0 and ACORDE). 
These detectors are used for event characterization, luminosity measurement
and for rejecting beam gas and pileup events.
The minimum bias trigger, comprising of V0 and ITS signals, is
$\sim$98\% efficient. Several triggers are planned to address topics which include 
selecting high multiplicity events and enriching the events with 
muons, electrons, photons and jets. 

\section{Physics performance}
Below we discuss some of the 
\mbox{p--p} physics topics, which can be accessed by ALICE based
on detailed studies of the detector performances using extensive simulations.

\noindent $\bullet$ {\bf Pseudorapidity and multiplicity distributions:} 
The basic measurements are those of the pseudorapidity and multiplicity distributions
of charged particles for minimum bias trigger conditions. 
The charged particle rapidity density as a function of $\sqrt{s}$ is expected to follow
Feynman scaling, the validity of which can be studied by ALICE.
The multiplicity distribution is expected to show strong departure from the KNO scaling,
which has been observed at lower CERN and Fermilab energies \cite{KNO,E735_0}. 
This departure has been attributed to multi-parton interactions and increasing contributions from jets.
These effects will play stronger roles at LHC energies. 
A high multiplicity trigger is employed to record events with large multiplicities for
the study of multi-parton interactions, parton saturation effects, 
correlations, rapidity gaps, HBT source sizes and event structures.
Long range forward-backward correlations will be
studied to understand the interplay of soft and hard physics
\cite{kittel}, whereas the charged-neutral correlations should shed light on the formation of
disoriented chiral condensates \cite{bjorken}. 
 
\noindent $\bullet$ {\bf \pT~spectra and $\ave{\pT}$-multiplicity correlations:} 
The measurements of \pT~spectra along with those of the pseudorapidity and multiplicity 
distributions are important for tuning model parameters in order to describe the bulk particle 
production mechanisms and underlying event structures. Proper description of \pT~spectra, both
at the high end where pQCD calculations should be applicable, and the low (non-perturbative)
region, is important for understanding any new emerging physics at LHC energies.
The advantages of the ALICE experiment is in its
capability for \pT~measurements over a wide range and in the
low-\pT~ cutoff. The correlation between $\ave{\pT}$ and multiplicity
provides information on the balance between particle production and transverse energy
of produced particles \cite{CDF,E735_1}. The correlation results will be studied as
functions of \pT~thresholds in order to get insight into the soft and hard sectors.

\noindent $\bullet$ {\bf Strangeness production:} 
Quantitative measures of strangeness are obtained by
the measurement of yield, spectra of various baryons and mesons and their
ratios, including those of resonances to stable particles. 
These studies help in characterizing the underlying event structure and
understanding the baryon production mechanism.
The energy dependence of strange particle
production exhibits a smooth behavior over a large range in c.m. energy, 
from tens of GeV to 1.8~TeV. At the same time, the correlation between the $\ave{\pT}$
of kaons and the charged-particle multiplicity at Tevatron energies 
has been observed to be stronger compared
to pions \cite{E735_1}. It will be important to study this behavior 
at LHC with access to much higher particle densities.

\noindent $\bullet$ {\bf Baryon number transfer in rapidity:}
In hadronic collisions, one of the open questions deals with the mechanism
by which baryon number (BN) gets transported to the central rapidity.
The standard mechanism 
of quark-diquark string breaking where the BN is carried by the valence quarks 
can not transport BN over a large range in rapidity. The concept of string
junctions has been introduced \cite{rossi} in which the valence quarks of the proton 
fragment independently, but are joined to a baryonic gluon field configuration.
If the BN is transfered dominantly
by gluons \cite{garvey} then the exchange probability is independent
of the rapidity interval, whereas in the string junction picture there will be a decrease
of this effect \cite{rossi}.  
Measurement of baryon distributions over a large rapidity interval (9.6 units) at LHC energies 
is expected to provide answers to different mechanisms of BN transport.
The BN asymmetry will be studied
for identified baryons (protons, $\Lambda$s, $\Xi$s and $\Omega$s) at low \pT~and 
as a function of particle multiplicity.

\noindent $\bullet$ {\bf Heavy flavour and quarkonia production:}
Measurement of heavy quark production (charm and bottom cross sections)
provide important tests of pQCD. Since the charm quarks
are created from initial gluon fusion, their measurements
at the new kinematical region of low Bjorken-{\it x} at LHC energies, would provide evidence for gluon shadowing.
ALICE will have simultaneous measurements in the
electronic channel in the central rapidity ($\pm 0.9$ units) 
and muonic channel in the forward regions (\mbox{2.5--4} in $\eta$).
The total charm cross section can be measured
with good accuracy because of high precision vertexing and close to zero \pT~cutoff. 
A high level trigger is being implemented using the transition radiation
detector (TRD) to enrich the electron sample. 

\noindent $\bullet$ {\bf Jets in \mbox{p--p}:}
The measurement of inclusive jet cross section provides the most stringent test
for pQCD predictions, and is essential for studying
the jet fragmentation and hadronizaton processes.
The jet studies in  \mbox{p--p} are essential 
for understanding the high density environment produced in heavy-ion
collisions. 
The jet reconstruction
in the initial runs will rely on the tracking 
measurements. The EMCAL
will add to the jet trigger capabilities,
extend the jet energy range, improve energy resolution, increase the efficiency and
reduce the bias on the jet fragmentation.
The tracking capabilities along with calorimetry represent an ideal tool for studies of jet structure modification
over a wide kinematic region of jet and associated particle momenta.

\noindent $\bullet$ {\bf Photon physics in \mbox{p--p}:}
Dominant contribution to direct photon production in \mbox{p--p} collisions
comes from gluon Compton and quark-antiquark annihilation processes. 
At large \pT, the photon production rate is proportional to the gluon distribution 
in the proton. At small \pT, the production rate is relevant for exploring the small $x(=2p_{\rm t}/\sqrt{s})$ 
region where next-to-leading order calculations become insufficient. High energy photons
will be used to tag charged jets emitted in the opposite direction in order to
study jet fragmentation functions.
The highly granular PHOS calorimeter will be able to identify photons over a large range
(\mbox{0.5--80}GeV) with excellent energy resolution (2\% above 3~GeV).
The ability to measure inclusive and direct photons down to
few GeV will allow ALICE to explore a new domain of QCD, 
well in the non-perturbative regime. 

\section{Summary}
ALICE has unique capabilities for \mbox{p--p} physics at the LHC energies
in terms of its low \pT~cutoff, excellent primary and secondary vertexing, particle identification over
a broad \pT~range and reconstruction of jets. These measurements are complementary 
to those of the dedicated \mbox{p--p} experiments.
Preparations for the data taking using the first 
\mbox{p--p} collisions are underway with hardware commissioning and 
offline readiness. LHC is scheduled to deliver colliding \mbox{p--p} beams at 
c.m. energy of 900~GeV in 2007, possibly followed by runs of 2.2~TeV, 5.5~TeV, leading to
14~TeV in 2008. This will allow for a systematic study of the \mbox{p--p} physics as function of
collision energy. 
ALICE will certainly play a major role as one enters into a new era of discovery 
with unprecedented energy and new kinematic regime offered by the LHC.

\end{document}